\documentstyle[preprint,eqsecnum,aps,pra]{revtex}

\begin{document}
\tightenlines
\renewcommand{\thefootnote}{\fnsymbol{footnote}}
\draft
\title{ Noise sustained pattern growth: bulk versus boundary effects}
\author {M.~L\"{u}cke\\}
\address{Institut f\"{u}r Theoretische Physik, Universit\"{a}t
des Saarlandes, D-66041~Saarbr\"{u}cken, Germany\\}
\author{A.~Szprynger\\}
\address{Institute of Low Temperature and Structure Research, Polish
Academy of Sciences, 50-950~Wroc\l aw 2, POB 937, Poland\\}
\renewcommand{\thefootnote}{\arabic{footnote}}
\setcounter{footnote}{0}
\date{\today}
\maketitle
\begin{abstract}
The effect of thermally generated bulk stochastic forces on the
statistical growth dynamics of forwards bifurcating propagating macroscopic
patterns
is compared with the influence of fluctuations at the boundary of a
semiinfinite system, $0<x$. To that end the linear complex Ginzburg-Landau
amplitude equation with additive stochastic forcing is solved by a
spatial Laplace transformation in the presence of arbitrary boundary
conditions for the fluctuations of the pattern amplitude at $x=0$.  
A situation where the latter are advected with an imposed through-flow from
an outside upstream part towards
the inlet boundary at $x=0$ is investigated in more detail. The spatiotemporal
growth behavior in the convectively unstable regime is compared with recent
work by Swift, Babcock, and Hohenberg where a special boundary condition
is imposed.

\end{abstract}

\narrowtext
\section{Introduction}\label{I}

Nonequilibrium extended systems that undergo a pattern forming
instability with nonzero group velocity \cite{CH-RevModPhys-93}  --- whether 
the latter arises in open-flow configurations
\cite{Kelly-94,Monkewitz} via an imposed 
through-flow 
\cite {MLK-Euro-89,MLK-PhysRevA-92,BAC-PhysRev-91,BCA-PhysD-92,BAC,TS-Euro-91,%
TS3-93,TS-PhysRev-91,TGS2-93,BLRS96,RBLMKS,OPML-Heat-95}
or internally from a bifurcation to travelling waves
\cite{DB-PhysLett-83,SR,RSTSHAB-PRL-91} ---
show a peculiar sensitivity to perturbations in the so called 
convectively unstable parameter regime \cite{BersBriggs,Huerre}. 
Therein initially spatially 
localized perturbations superimposed on some basic unstable, macroscopically 
unstructured nonequilibrium state of the system are 
advected downstream by the group velocity. Simultaneously they 
are selectively amplified \cite{Huerre,Deissler}  according to the 
deterministic amplification mechanism that occurs near the 
pattern forming bifurcation \cite{CH-RevModPhys-93}. Thus, advection spatially 
deconvolutes the pattern growth from its cause -- the perturbation source.
But the latter leaves its signature via the bifurcation-induced, 
deterministic amplification process 
on the spatiotemporal properties of the downstream 
growing macroscopic nonequilibrium structures. The statistical dynamics
of fluctuations of these structures is easily detectable 
\cite {MLK-PhysRevA-92,BAC-PhysRev-91,BCA-PhysD-92,BAC,TS3-93,%
TS-PhysRev-91,TGS2-93,SR,RSTSHAB-PRL-91} thus 
providing information about the small perturbations that have triggered
their growth. 

The feasibility to 
distinguish the influence of different perturbation sources -- e.g. those that
are external to the system under study versus internal thermal noise -- 
on the macroscopic pattern growth in a very direct way has attracted
much research activity recently \cite {BAC-PhysRev-91,BCA-PhysD-92,BAC,TS3-93,%
TS-PhysRev-91,TGS2-93,SR,RSTSHAB-PRL-91}. Here one has to keep 
in mind that the
deterministic bifurcation dynamics of the macroscopic structure forming
instability determines which of the modes that a perturbation source 
emits is amplified and which one decays in downstream direction. And as
long as the mode amplitudes are still small each of the perturbation modes
evolves separately and independently: the bifurcation dynamics
is initially linear with saturating nonlinearities being irrelevant 
thus producing exponential growth with time and downstream distance.

Hence, in a system with group velocity in positive $x$ direction one expects 
that fluctuations generated near the inlet boundary at $x=0$
and/or perturbations that might enter the system via the inlet are most
important.
They have the longest advection/amplification time to grow in the flow before
they reach a particular observation point at a downstream location $x>0$.
Therefore a careful analysis of the influence of 
the inlet boundary condition relative to the effect of 
bulk (fluctuating) forces is necessary in 
particular in view of the fact that both sources 
cause exponential growth. Such an analysis 
should be able to shed some light on the 
problem whether in bounded experimental 
systems it is internal bulk generated thermal noise 
that triggers and sustains the growth of 
macroscopic vortex structures in the 
convectively unstable regime or other kind of 
environment-produced perturbations that might, 
e. g., be advected from the inlet boundary into 
the bulk. 

On the theoretical side the effect of noise on the 
pattern forming process has been investigated 
mainly within the framework of the complex 
Ginzburg-Landau amplitude equation (CGLE). It 
describes the slow spatio-temporal behavior of 
the complex amplitude $A(x,t)$ of the critical 
mode of the pattern that can grow first when 
crossing the bifurcation threshold. Bulk thermal 
noise is incorporated in this approximation 
scheme by fluctuating volume forces $f(x,t)$ that 
act additively on $A(x,t)$ and that are derived 
\cite{Graham,H-S92,SZ,SBH,Treiber96}
for local thermal equilibrium conditions of driven systems
from the linear Landau-Lifshitz equations of 
fluctuating hydrodynamics around global thermal equilibrium \cite{LanLif}. 
But the problem 
of the inlet boundary condition that 
arises with the CGLE in a finite domain has 
been addressed only in some idealized way: 
Deissler \cite{Deissler94} imposes $A = 0$ at the inlet.
L\"{u}cke and Recktenwald \cite{LR} investigate the pattern
growth resulting from 
an inlet boundary condition that is produced by 
thermal equilibrium transverse momentum 
fluctuations which are advected into the system 
at $x = 0$. Babcock, Ahlers, and Cannell \cite{BAC} have solved the
stochastic CGLE numerically in a system that was augmented in upstream direction,
i.e., to negative $x$ by a small subcritical part and compared with imposing
$A=0$ at $x=0$. Swift, 
Babcock, and Hohenberg (SBH) \cite{SBH}  consider 
solutions of the stochastic CGLE where the 
amplitude fluctuations $A(x=0,t)$ at the boundary 
are strongly correlated with the bulk thermal 
forcing $f$.

In this paper we first of all solve the stochastic CGLE subject to
arbitrary inlet boundary conditions with a spatial Laplace
transformation method. We investigate the influence of  thermally
generated bulk additive forcing on the statistical dynamics of
downstream growing amplitude fluctuations in comparison with the
effect of statistically stationary inlet fluctuations that are
independent of the bulk forcing. Furthermore, we use a simple model to
estimate fluctuations that are swept with an imposed through-flow
into some experimental systems via the inlet. This allows to assess
more quantitatively the importance of such inlet fluctuations 
relative to bulk thermal forcing. Experimental setups to tune and to
control the former are shortly discussed as well. Finally we compare
in detail with the results of SBH. 

Our paper is organized as
follows: In Sec.~\ref{II} the linear stochastic CGLE with thermally  
generated bulk additive forcing is briefly reviewed. The solution for
the fluctuating amplitudes in frequency space, $A(x, \omega)$, and
their correlation spectra are given in Sec.~\ref{III}. The next section
presents the model for inlet fluctuations and compares their
downstream evolution with the growth of bulk generated fluctuations.
The last section contains a summary. In Appendix \ref{appa} we present the
"space-retarded, causal" Green function of our solution in the
convectively unstable regime. In Appendix \ref{appb} we compare our results
for Green function, amplitude fluctuations, and correlation spectra
with those of SBH.

\section{Systems}\label{II}

The systems that we have in mind show a continuous,
i.e., hysteresis-free transition, say, from a spatially 
homogeneous state to a spatially structured one. Thus, a 
spatially periodic solution branches in a forwards bifurcation
out of the basic solution at a critical control parameter. 
The most intensively studied examples for such systems 
are the Rayleigh-B\'enard (RB) problem of straight roll 
convection vortices in a fluid layer heated from below 
\cite{CH-RevModPhys-93} or
the rotating Taylor-Couette (TC) problem of axisymmetric
toroidal Taylor vortices in the annulus between concentric
cylinders of which the inner one is rotating 
\cite{CH-RevModPhys-93,TAGG-94}. But our
investigation being based on the CGLE for such 1D patterns also 
includes other problems for which this equation captures 
the essential parts of the pattern formation process.
Nevertheless we shall use in the remainder of this paper the 
terminology of the RB or TC system.
Thus, we consider in particular open RB (TC) setups with 
an externally imposed horizontal (axial) mean flow through
a not too wide rectangular convective channel (annulus
between concentric cylinders). Then the deterministic
field equations show an oscillatory instability. The structureless,
stationary basic flow without vortices is stable for Rayleigh
numbers $Ra$ (Taylor numbers $T$) below a critical threshold
$Ra_c (Re)\, [T_c (Re)]$ that depends on the through-flow 
Reynolds number $Re$. At the critical threshold a stable, 
spatially extended, temporally oscillatory state of downstream 
propagating vortices branches off the basic state. The latter is
unstable and the former is stable for supercritical driving.

We use the relative control parameter
\begin{equation}
\mu = \frac{Ra}{Ra_c \left(Re \right)} - 1 \quad \mbox{or} \quad
\mu = \frac{T}{T_c \left(Re \right)} - 1
\label{MU}   
\end{equation}
to measure the distance from the onset of propagating vortex flow for
$Re \neq 0$ and of stationary vortex flow for $Re=0$, 
respectively. In
this notation $\mu_c=0$, i.e.,
\begin{equation}
\epsilon_c \left(Re \right) =  \frac{Ra_c \left(Re \right)}{Ra_c \left(Re=0
\right)} -1 \quad \mbox{or} \quad 
\epsilon_c \left(Re \right) =  \frac{T_c \left(Re \right)}{T_c \left(Re=0
\right)} -1 
\label{EPSILON_C} 
\end{equation}
is the critical reduced threshold for onset of  propagating vortex flow.
The shear forces of the through-flow slightly stabilize
the homogeneous basic state, so $\epsilon_c \left(Re \right)$
slightly increases with $Re$
\cite{Kelly-94,MLK-Euro-89,BAC-PhysRev-91,BCA-PhysD-92,BAC,%
RLM-PhysRevE-93}. Note that in the RB system we consider so-called
transverse vortices whose roll axes are enforced to lie
perpendicular to the through-flow direction by the
geometrical design of the rectangular convection channel. So for not too
large $Re$ the critical wave vector of the vortex patterns is oriented
in both cases along the $x$-direction of the through-flow, 
${\bf k}_c = k_c {\bf e}_x$.

\subsection{Amplitude equation approximation}\label{IIA}

For small supercritical control parameters, $0 < \mu \ll 1$, the macroscopic 
vortex dynamics is governed by a narrow band of near critical modes. This
allows to represent the hydrodynamic fields within an amplitude equation
approximation \cite{CH-RevModPhys-93} by a complex amplitude $A(x,t)$ multiplying 
the critical mode
\begin{equation}
w\left(x,{\bf r}_{\perp};t\right) = A\left(x,t\right) e^{i
\left(k_c x - \omega_c t\right)} \hat{w}\left({\bf r}_{\perp}\right) + c.c.
\label{WAMPEQ} 
\end{equation}
Here $w$ is for example a component of the vortex velocity field.
The critical wave number $k_c$, frequency $\omega_c$ 
\cite{MLK-Euro-89,BAC-PhysRev-91,BCA-PhysD-92,%
BAC,RLM-PhysRevE-93,CA-FluidMech-77,TJ-FluidMech-81}, and eigenfunction
$\hat{w}({\bf r}_{\perp})$ \cite{CA-FluidMech-77,AR-Dipl-91} depending on the
coordinates ${\bf r}_{\perp}$ perpendicular to the critical wavevector
of the downstream propagating vortex structures are known as functions of 
 $Re$ from a linear stability analysis of the basic flow.

 In the absence of
fluctuations the spatiotemporal behavior of the complex vortex amplitude 
$A(x,t)$ is determined by the 1D CGLE
\begin{equation}
\tau_0 (\partial_t + v \partial_x)A (x,t) =
\left[(1+ic_0)\mu + (1+ic_1)\xi_0^2\partial_x^2\right] A (x,t) 
+\mbox{nonlinear terms}.
\label{GLE_DET} 
\end{equation}
The coefficients $\tau_0$ and $\xi^2_0$ are even in $Re$ while the 
group velocity $v$ and the
imaginary parts $c_0$, $c_1$ are odd in $Re$ 
\cite{MLK-PhysRevA-92,RLM-PhysRevE-93}. 

\subsection{Absolute and convective instability}\label{IIB}

For small $\mu$ and $Re$ the control parameter plane is divided
into three domains
characterized by different growth behavior of {\it linear}
perturbations of the basic flow state.  
In the presence of through-flow 
one has to distinguish \cite{BersBriggs}
between the spatiotemporal growth behavior of spatially localized
perturbations and of spatially extended ones. 
Below the threshold $\mu=0$ 
for onset of propagating vortex patterns any 
perturbation, spatially localized as well as extended, decays. This 
is the parameter regime of absolute stability of the basic state.

For $\mu > 0$ extended perturbations can grow.  However, a spatially localized
perturbation, i.e., a packet of plane wave perturbations is 
advected in the so-called convectively unstable parameter regime 
faster downstream than it grows --- while growing in the comoving 
frame it moves out of the system
\cite{BersBriggs,Huerre,Deissler,TES-PhysRev-90}. Thus, the 
downstream as well as the upstream facing intensity front of the vortex
packet move into the same direction, namely, downstream. 
A spatially localized perturbation 
is blown out of any system of finite length and the basic 
state is reestablished.
It therefore requires a persistent source of perturbations like, e.~g., noise 
operating in the bulk or fluctuations entering the system via the
inlet to sustain a vortex pattern in the convectively unstable regime
\cite{Deissler}. The  spatiotemporal 
behavior of such "noise sustained structures" \cite{BAC,TGS2-93} being
determined by the noise source thus
allows to infer properties of the latter from the former.

In the absolutely unstable regime a localized
perturbation grows not only in the downstream direction but it grows and
spatially expands also in the upstream direction until the upstream
propagating front encounters the inlet in a finite system
\cite{BLRS96}. 
The final nonlinear stucture 
\cite{MLK-Euro-89,MLK-PhysRevA-92,TS-Euro-91,TS3-93,BLRS96,RBLMKS,%
OPML-Heat-95,TS1-93}
resulting in such a situation is self sustained and stable. For sufficiently
large amplitudes this structure is dominated by the deterministic 
contributions in the relevant balance equations and thus it is insensitive to,
say, thermal noise.

Within the framework of the
amplitude equation the boundary between absolute and convective
instability is given by \cite{Deissler}
\begin{equation}
 \mu^{c}_{conv} = \frac{\tau^2_0 v^2}{4 \left(1+c^2_1\right)\xi^2_0}.
\label{MUCCONV}
\end{equation}
Thus, the convectively unstable regime investigated here
is characterized by $0< \mu < \mu^{c}_{conv}$ or, equivalently, by the
reduced group velocity 
\begin{equation}
V = \frac{\tau_0}{\xi_0 \sqrt{\left(1+c^2_1\right)\mu}}
v = 2\sqrt{\frac{\mu^{c}_{conv}}{\mu}}
\label{V}
\end{equation}
being larger than 2.
It should be mentioned that the amplitude equation  approximation
(\ref{MUCCONV}) 
of $\mu^{c}_{conv}$ describes the boundary between absolute and
convective instability resulting from the full hydrodynamic field equations 
\cite{BAC,RDprivate}
very well in the RB and TC systems for the small Reynolds numbers considered
here.

\subsection{Stochastic amplitude equation}\label{IIC}

The starting point of most theoretical approaches for describing the
effect of thermal noise on macroscopic hydrodynamic pattern growth is
the Langevin concept of Landau and Lifshitz \cite{LanLif}. Therein the linear
hydrodynamic field equations are heuristically supplemented by Gaussian
distributed statistically stationary stochastic forces with zero mean.
Their two-point correlation functions reflect the isotropy and
translational invariance of an unbounded system and their frequency and
wave vector spectra are white. Projection \cite{Graham,H-S92,SZ,SBH,Treiber96}
of these 
equations onto the critical modes leads then to the linear stochastic CGLE
\begin{equation}
\left[ \tau_0 (\partial_t + v \partial_x) - (1+ic_0)\mu
-(1+ic_1)\xi_0^2 \partial_x^2\right] A (x,t) = f(x,t)
\label{GLE_Ax}
\end{equation}
with Gaussian distributed complex forces $f(x,t)$ that are
$\delta$-correlated in space and time
\begin{mathletters}
\begin{eqnarray}
< f (x,t) > =  < f (x,t) f (x',t') > = 0, \\
< f (x,t) \left[ f (x',t') \right]^* > = 
 F \delta (x-x') \delta (t-t').
\end{eqnarray}
\label{f_corr}
\end{mathletters}
Here the star implies complex conjugation.
The noise strength $F$ has been evaluated by several authors
\cite{Graham,H-S92,SZ,SBH,Treiber96} 
for the RB and TC systems without through-flow and with different
boundary conditions. Since we do not need the specific values of $F$
we shall not dwell on questions whether and how the forces are changed 
by finite-size effects, system geometry, and the through-flow. We
rather take eqs. (\ref{GLE_Ax}, \ref{f_corr}) as the starting point for
our investigations. 

However, for later use we mention that our $F$ is related
via $F = 2\xi_0\tau_0 F_A^{SBH}$ to the noise strength $F_A^{SBH}$ given
in eqs. (2.10 - 2.16) of SBH. We should also like to mention that the 
value of $F$ characterizing a particular system depends on the arbitrary
choice of the normalization condition for the linear critical mode 
$\hat{w}\left({\bf r}_{\perp}\right)$ that enters into Eq. (\ref{WAMPEQ}) 
and that establishes the relation between the physically observable
order parameter field $w\left(x,{\bf r}_{\perp};t\right)$ and the complex
amplitude $A$ of the critical mode.


\section{Linear stochastic CGLE in semiinfinite systems} \label{III} 
 
In this section we discuss the statistical dynamics of pattern growth in the 
convectively unstable parameter regime as described by the  
stochastically forced linear CGLE ~(\ref{GLE_Ax}). We consider a 
semiinfinite system with the through-flow entering  
the system at $x=0$. This flow transports fluctuations into 
the system that we specify by imposing an inlet boundary condition  
$A\left(x=0,t\right)$ 
on the amplitude $A$. We consider situations in which $A\left(x=0,t\right)$ 
 is the result 
of a statistically stationary process; see Sec. \ref{IVA} for a model. Thus, 
with the bulk forces $f(x,t)$ being also statistically stationary we  
restrict our investigations to the long-time statistically stationary  
fluctuations of $A(x,t)$ --- initial transients occurring in experiments after, 
e.g. changing control parameters have died out. Under these circumstances we 
found it convenient to perform part of our analysis in frequency space. 
After a temporal Fourier transformation 
\begin{equation} 
A(x,\omega) = \int _{-\infty}^{\infty} dt e^{i \omega t} A(x,t) 
\end{equation} 
the stochastic CGLE becomes an ordinary, second order, inhomogeneous  
differential equation  
\begin{mathletters} 
\label{GLE} 
\begin{equation}
\left[ \tau_0 (- i \omega + v \partial_x) - (1+ic_0)\mu 
 - (1+ic_1)\xi_0^2\partial_x^2\right] A (x,\omega) = f (x,\omega) 
\end{equation}
that is to be solved subject to the inlet boundary condition  
$A(x=0,\omega)$. A condition on the derivative of $A$ at the  
inlet or, equivalently, on the behavior of $A$ for $x\rightarrow +\infty$  
follows from a physical requirement discussed in Sec. \ref{IIIA} below.

For later use we rwrite the CGLE in the form
\begin{equation} 
- (1+ic_1)\xi_0^2 (\partial_x - i K_1)(\partial_x -  
i K_2) A (x, \omega) = f (x,\omega) 
\end{equation} 
\end{mathletters} 
that displays the spatial eigenvalues $K_{1,2}$ of the CGLE explicitly ---
in the solution of Eq. (\ref{GLE}) there appear contributions of the form 
$e^{iK_1x}$ and $e^{iK_2 x}$ where $K_{1,2}$ are the complex $\omega$-dependent 
characteristic exponents, i.e., spatial eigenvalues of Eq. (\ref{GLE}) 
\begin{equation} 
\xi_0 \sqrt{1+ic_1}K_{1 \atop 2} = \pm i \sqrt{(1-ic_1)\mu^{c}_{conv} -  
(1+ic_0)\mu - i \omega \tau_0}- i \sqrt{(1-ic_1)\mu^{c}_{conv}}. 
\label{eigenvalues} 
\end{equation}  
The physically relevant characteristic exponent that controls the  
spatial growth of vortices in downstream direction is $K_1$ 
whereas the second eigenvalue of the CGLE is such that $Im K_2<0$ for 
all $\mu$ and all $\omega$. 
  
$Im K_1(\mu;\omega)$ is for $\mu<0$ positive 
so that the spatial growth rate $-Im K_1$ is negative implying spatial  
decay of each frequency mode $\omega$. For $\mu>0$ selective amplification 
of frequency modes out of a finite band $\omega_-< \omega < \omega_+$  
occurs   
\begin{mathletters} 
\begin{eqnarray} 
Im K_1(\omega;0<\mu<\mu^{c}_{conv}) < 0   
\qquad\mbox{ for~ $\omega_-< \omega <\omega_+$ } \\ 
Im K_1(\omega;0<\mu<\mu^{c}_{conv}) > 0  
\qquad\mbox{ for~ $\omega<\omega_-$ and $\omega>\omega_+$. }  
\end{eqnarray} 
\end{mathletters} 
The band limits for amplification are 
\begin{equation} 
\tau_0 \omega_{\pm} = 
 \pm 2 \sqrt{(1+c_1^2)\mu\mu^{c}_{conv}} + (c_1-c_0)\mu.  
\label{omega_b} 
\end{equation} 
$Im K_1$ has a quadratic minimum at the frequency 
\begin{equation}  
\omega_m=-(c_0+c_1)\mu/\tau_0 
\end{equation} 
 which is, due to the smallness of 
$c_0$ and $c_1$, in general located close to $\omega=0$. Thus, the frequency 
mode with largest spatial growth is $\omega_m$. 
 
For plots of $-Im K_1$ versus $\omega$ see, e.g., \cite[Fig.~1]{LR} and 
\cite[Fig.~14]{BAC}. The relations to our exponent $K_1$ are  
$i K_1 = \kappa \cite{LR} = \beta\cite{BAC}$. 
 
\subsection{Spatial Laplace transformation}\label{IIIA} 
 
We have solved the CGLE (\ref{GLE}) with the method of Green's functions 
and, equivalently, using a spatial Laplace transformation. The latter method 
most clearly displays the influence of the inlet boundary properties on the 
solution. Thus, we start with the latter method here and 
present the relation to the former in Appendix \ref{appa}. 
 
We use the spatial Laplace transformation 
\begin{equation} 
A(K)= -i \int_0^{\infty} dx e^{-iKx} A(x) \,\,\, ; \,\,\,K \,\,\, 
\mbox{complex}. 
\label{Ltdef} 
\end{equation} 
Here the second argument -- $\omega$ or $t$ -- of $A$ is suppressed. Using 
the transformation properties 
\begin{mathletters} 
\begin{eqnarray} 
\partial_x A(x) &\longleftrightarrow & i K A(K) + i A(x)\vert _{x=0},  
\\ 
\partial_x^2 A(x) &\longleftrightarrow & -K^2 A(K) +i ( \partial_x  
 + i K) A(x) \vert _{x=0} 
\end{eqnarray} 
\end{mathletters} 
the Laplace transformed CGLE (\ref{GLE}) reads 
\begin{eqnarray} 
(1+ic_1)\xi_0^2 (K - K_1) (K - K_2) A (K, \omega) = 
\nonumber\\ 
i(1+ic_1)\xi_0^2 \left[ \partial_x + i (K - K_1 - K_2) \right]A 
(x, \omega) \mid_{x=0} + f (K, \omega). 
\end{eqnarray} 
Its mathematically formal, {\em physically unrestricted} solution 
\begin{equation} 
A (K, \omega) = \frac{i \left[ \partial_x + i (K - K_1 - 
K_2) \right] A (x, \omega)\mid_{x=0}}{(K - K_1) (K - K_2)} + 
\frac{f (K,\omega)}{(1+ic_1) \xi_0^2 (K - K_1)(K - K_2)} 
\label{A(K,omega)} 
\end{equation} 
contains poles in the complex $K$-plane at $K_1$ and at $K_2$. 
 
Now in the {\em physical} solution the residue of the pole at $K_2$ has to  
vanish, i.e., 
\begin{equation} 
\lim_{K \to K_2} \left[ (K - K_2) A (K, \omega) \right] = 0. 
\label{generalrestriction} 
\end{equation} 
Otherwise there would be a contribution ~\,$\sim$\, $e^{iK_2 x}$ for  
$x\to +\infty$ giving rise to spatial growth and a divergence of  
$A(x\to +\infty, \omega)$ even for subcritical driving $\mu<0$. 
The physically motivated requirement that there is no growth  
~\,$\sim$\, $e^{iK_2 x}$ for $x \to +\infty$ is equivalent to selecting  
the "retarded" or causal Green function in the solution of the CGLE  
(cf. Appendices \ref{appa} and \ref{appb1}). Therein the solution is 
represented by a  
superposition of "waves" which propagate (in the frame comoving with 
velocity $v$) outwards, i.e., away from a localized perturbation 
source ~\,$\sim$\, $\delta(x-x_0) \delta(t-t_0)$ that creates a   
perturbation pulse at $x_0$, $t_0$. For $\mu<0$ the "waves" are damped while 
for $0<\mu<\mu^{c}_{conv}$ they grow in downstream direction within 
the frequency band (\ref{omega_b}). 
On the other hand, the "advanced" Green function 
describes inwards propagating waves 
 ~\,$\sim$\, $e^{i[K_2(\omega) x - \omega t]}$ 
that create the pulse at $x_0$, $t_0$. They would cause 
even for $\mu<0$ a divergence of $A$ at $x \to +\infty$. 
 
Note that the requirement (\ref{generalrestriction}) implies that for a given 
forcing $f(x, \omega)$ the initial slope $\partial_x A(x, \omega)\mid_{x = 0}$ 
of the amplitude at the inlet is fixed by the inlet value $A(x=0, \omega)$ 
and the forcing $f(K_2,\omega)= -i \int_0^{\infty} dx e^{-iK_2x} f(x,\omega)$. 
From eqs. (\ref{A(K,omega)}) and (\ref{generalrestriction}) one obtains the 
relation 
\begin{equation} 
i \left( \partial_x - i K_1 \right) A(x, \omega)\mid_{x = 0} 
+ \frac{f (K_2, \omega)}{(1+ic_1)\xi_0^2}=0. 
\label{restriction} 
\end{equation} 
Using this restriction to eliminate the slope in (\ref{A(K,omega)}) the 
physical solution of the CGLE can be written as the sum of two terms 
\begin{mathletters} 
\label{A(K,omega)restricted} 
\begin{equation} 
A (K, \omega) = A_{in}(K, \omega) + A_b (K, \omega).   
\end{equation} 
Here  
\begin{equation} 
\label{Ain(K,omega)} 
 A_{in}(K, \omega)= - \frac{A (x=0, \omega)}{K - K_1} 
\end{equation} 
is the  contribution from the inlet fluctuations $A (x=0, \omega)$ while  
\begin{equation} 
\label{Ab(K,omega)} 
A_b (K,\omega) =  
\frac{f(K,\omega) - f (K_2,\omega)}{(1+ic_1) \xi_0^2 (K -K_1)(K - K_2)} 
\end{equation} 
\end{mathletters} 
is caused by bulk forcing. In real space the solution reads 
\begin{mathletters} 
\label{A(x,omega)restricted} 
\begin{equation} 
A (x, \omega) = A_{in}(x, \omega) + A_b (x, \omega)   
\end{equation} 
where  
\begin{equation} 
\label{Ain(x,omega)} 
 A_{in}(x, \omega)=  A (x = 0, \omega) e^{i K_1 x}  
\end{equation} 
and  
\begin{eqnarray} 
\label{Ab(x,omega)} 
A_b (x,\omega) =  
\frac{i}{(1+ic_1)\xi_0^2 (K_1 - K_2)}\int_0^x dx' 
\left[ e^{i K_1 (x - x')} - e^{i K_2 (x- x')} \right] f (x',\omega) 
\nonumber\\ 
- \frac{i}{(1+ic_1)\xi_0^2 (K_1 - K_2)}\left[ e^{i K_1 x} - e^{i K_2 x} 
\right]\int _0^\infty  dx' e^{-i K_2 x'} f (x',\omega). 
\end{eqnarray} 
\end{mathletters} 
Note that the bulk part vanishes at the inlet boundary, 
$A_b (x=0,\omega) = 0$. 
 
\subsection{Digression: constant bulk and inlet forcing} 
 
In this subsection we make a digression to investigate the special case  
\begin{equation} 
f(x, \omega) = 2\pi\delta(\omega)f_0 \, ;\qquad 
A(x=0, \omega)=2\pi\delta(\omega) A_0  
\end{equation} 
of constant (deterministic) forcing $f(x,t) = f_0$ in the bulk and constant  
inlet boundary condition $A(x=0,t) = A_0$. This investigation is useful 
and instructive for two reasons: (i) It allows to check the general solution 
(\ref{A(x,omega)restricted}) for a special case that can easily be solved 
analytically in a direct way without Laplace transformation and that 
furthermore can easily be solved numerically. (ii) It gives insight into 
the way how initial slope  $\partial_x  A\mid_{x = 0}$, inlet value   
$A_0$, and bulk forcing $f_0$ are tied to each other. 
 
We consider in this subsection the stationary solution $A(x)$ of the CGLE 
, i.e., $A(x,\omega)=2\pi\delta(\omega) A(x)$.  
Then (\ref{Ain(x,omega)}) gives 
\begin{mathletters} 
\begin{equation} 
\label{Ain(x,0)} 
A_{in}(x) =  A_0 e^{i K_1^0 x}  
\end{equation} 
and (\ref{Ab(x,omega)}) becomes  
\begin{equation} 
\label{Ab(x,0)} 
A_b (x) =  
\frac{f_0}{(1+ic_1)\xi_0^2 K_1^0 K_2^0} \left(1- e^{i K_1^0 x} \right)  
\end{equation} 
with  
$ K_1^0=K_1(\omega=0)$ and $ K_2^0=K_2(\omega=0)$ given by 
(\ref{eigenvalues}). The complete solution reads 
\begin{eqnarray}  
\label{A(x,0)} 
A(x) &=& A_{in}(x)+A_b (x) 
\nonumber \\  
&=& A_0 e^{i K_1^0 x} - \frac{f_0}{(1+ic_0)\mu} \left(1- e^{i K_1^0 x} \right).  
\end{eqnarray} 
\end{mathletters} 
Here we have used $(1+ic_1)\xi_0^2 K_1^0 K_2^0=-(1+ic_0)\mu$. The physically 
motivated restriction (\ref{generalrestriction}, \ref{restriction}) on the  
slope at the inlet becomes in our special case  
\begin{eqnarray} 
\partial_x A(x) \mid_{x = 0} = i K_1^0 A_0 -i \frac{f_0} 
{(1+ic_1)\xi_0^2 K_2^0} 
\nonumber \\  
=i K_1^0 \left[ A_0 + \frac{f_0}{(1+ic_0)\mu} \right] 
\end{eqnarray} 
since $f(K_2,\omega) = - f_0 2\pi\delta(\omega)/K_2^0$ for constant 
forcing. We have checked the validity of (\ref{A(x,0)}) also by direct 
numerical integration of the (nonlinear) CGLE.

\subsection{Correlation functions} 
 
Here we investigate the frequency spectrum  
\begin{equation} 
C(x,\omega)= 
\int_{-\infty}^\infty  d(t-t') e^{i \omega (t-t')} \, C(x,\mid t - t' \mid ) 
\end{equation} 
of the time-displaced autocorrelation function 
\begin{equation} 
\label{C(x,t)} 
C(x,\mid t - t' \mid )= <A (x,t) \left[A (x,t') \right]^* >  
\end{equation} 
of amplitude fluctuations $A(x,t)$ at the same downstream position $x$. 
Since the semiinfinite system with through-flow is spatially not  
translational invariant the correlations (\ref{C(x,t)}) depend 
explicitly on the position $x$. On the other hand, the statistical 
stationarity of the forcing processes in the bulk and at the inlet causes 
a temporal dependence only via $\mid t - t' \mid$. Using this property 
one has 
\begin{equation} 
C(x,\omega) = \frac{< A(x,\omega) \left[ A (x,\omega') \right]^* >} 
{2\pi\delta(\omega-\omega')}. 
\end{equation} 
 
We consider inlet fluctuations $A(x=0,\omega)$ that are determined  
by forces operating outside the bulk fluid volume, for instance by  
fluctuating forces in the upstream part of an experimental setup. Thus, in 
the convectively unstable regime the 
contribution $A_b(x,\omega)$ (\ref{Ab(x,omega)}) due to bulk forcing is  
uncorrelated with the inlet fluctuations $A(x=0,\omega)$ so that    
\begin{equation} 
\label{<A(x=0)Ab(x)>} 
< A(x=0,\omega) \left[ A_b (x,\omega') \right]^* > = 0 =  
< A_{in}(x,\omega) \left[ A_b (x,\omega') \right]^* >.  
\end{equation} 
Therefore  
\begin{mathletters} 
\label{C(x,omega)}  
\begin{equation} 
C(x,\omega) = C_{in}(x,\omega) + C_b(x,\omega)  
\end{equation} 
is the sum of correlations  
\begin{equation} 
\label{Cin(x,omega)} 
 C_{in}(x,\omega) = e^{-2 Im K_1 x} D(\omega) 
\end{equation} 
of the contribution $A_{in}(x,\omega)$ (\ref{Ain(x,omega)}) caused 
by inlet fluctuations and of correlations 
\begin{eqnarray} 
\label{Cb(x,omega)} 
C_b(x,\omega) =  \frac{F}{2 (1+c_1^2) \xi_0^4  
\mid K_1 - K_2 \mid^2 } \Biggl\{ \frac{1}{Im K_1} - \frac{1}{Im K_2}  
\nonumber\\ 
  +\left( 4 Im \frac{1} {K_1^* - K_2} - \frac{1}{Im K_1} 
- \frac{1}{Im K_2} \right) e^{-2 Im K_1 x}  
\nonumber\\ 
+ 4 Re \left[ \left( \frac{1}{2 Im K_2} +  
\frac{i}{K_1^* - K_2} \right) e^{i (K_1-K_2) x} \right] \Biggr\} 
\end{eqnarray} 
\end{mathletters} 
of $A_b(x,\omega)$ (\ref{Ab(x,omega)}) caused by bulk forcing. 
Here  
\begin{equation} 
\label{D(omega)}  
D(\omega) = C_{in}(x=0,\omega) 
=\frac{< A(x=0,\omega) \left[ A (x=0,\omega') \right]^* >} 
{2\pi\delta(\omega-\omega')} 
\end{equation} 
is the fluctuation spectrum of the inlet amplitude. The expression 
(\ref{Cb(x,omega)}) for  
\begin{equation} 
C_b(x,\omega) = \frac{< A_b(x,\omega) \left[ A_b(x,\omega') \right]^* >} 
{2\pi\delta(\omega-\omega')}. 
\end{equation} 
is obtained directly from (\ref{Ab(x,omega)}) by using the force correlations 
\begin{equation} 
< f (x, \omega ) \left[ f (x', \omega ') \right]^* > = \delta (x-x') 
2 \pi \delta (\omega - \omega ') F. 
\end{equation} 
 
Let us briefly discuss the small-$x$ and the large-$x$ behavior 
of $C(x,\omega)$: Since $C_b(x=0,\omega)=0$ one has  
$C(x=0,\omega)=C_{in}(x=0,\omega)=D(\omega)$. Furthermore, $C_{in}$ 
varies linearly and $C_b$ quadratically in $x$ close to the inlet. 
 
The large-x behavior is physically more interesting since it tells how 
and which frequency modes are amplified. Here one has to distinguish 
the two cases (i) $Im K_1>0$  and (ii) $Im K_1<0$. The case (i) holds for 
all $\omega$  if $\mu<0$ and for $\omega$ outside the band 
($\omega_-,\omega_+$) of growing frequency modes if $\mu>0$. For these damped 
modes the autocorrelation approaches at large distances from the inlet the  
limit value  
\begin{mathletters} 
\label{Cinf} 
\begin{equation} 
\lim_{x\to \infty} C(x,\omega) = 
\frac{F}{2 (1+c_1^2)\xi_0^4 \mid K_1 - K_2 \mid^2 } 
\left(\frac{1}{Im K_1} - \frac{1}{Im K_2} \right).  
\label{Cx_inf} 
\end{equation} 
The fluctuations of these damped modes are influenced and sustained at large 
distances from the  inlet only by the bulk forcing. The  
limiting  value (\ref{Cx_inf}) coincides with the autocorrelation 
\begin{equation} 
\label{Cinf(omega)} 
C_\infty (\omega) =  
\int_{- \infty}^\infty \frac{d k}{2 \pi} 
\frac{F} {\mid (1+ic_1)\xi_0^2  k^2 - \mu (1+ic_0) - i(\omega - v k)  
\tau_0 \mid^2} 
\end{equation} 
\end{mathletters} 
of amplitude fluctuations in an infinite system, 
$- \infty<x<\infty$, that is translational invariant without inlet boundary 
conditions. 
 
In case (ii) of $Im K_1<0$, ~i.e., for $\mu>0$ and  
$\omega_-< \omega<\omega_+$  there is exponential growth. Then the  
correlation function is dominated for large $x$ by   
\begin{equation} 
 C(x\rightarrow \infty,\omega) \rightarrow    
\left[ D(\omega) 
- \frac{F}{2 (1+c_1^2) \xi_0^4 \mid K_1^* - K_2 \mid^2 }   
\left(\frac{1}{Im K_1} + \frac{1}{Im K_2} \right) \right]e^{-2 Im K_1 x}. 
\label{large_x}  
\end{equation} 
To derive ~(\ref{large_x}) from ~(\ref{C(x,omega)}) we have used the relation 
\begin{equation} 
\frac{1}{\mid K_1-K_2\mid^2} \left( 4Im \frac{1}{K_1^*-K_2} - \frac{1} 
{Im K_1} - \frac{1}{Im K_2}\right) = \frac{-1}{\mid K_1^*-K_2\mid^2}\left( 
\frac{1}{Im K_1} + \frac{1}{Im K_2}\right). 
\label{identity}  
\end{equation} 
 
Eqs. (\ref{C(x,omega)}) for the spectra of amplitude fluctuations and their 
large-$x$ limiting behavior are the central result of this 
paper. It allows to assess the influence of (externally generated) inlet 
fluctuations and of bulk stochastic forces separately. 


\section{Inlet and bulk generated fluctuations}\label{IV}

Here we first introduce a simple model for the frequency spectrum 
$D(\omega)$ (\ref{D(omega)}) of inlet fluctuations of the amplitude
$A(x=0)$. It allows to make quantitative estimates of the importance of 
their contributions $C_{in}(x,\omega)$ (\ref{Cin(x,omega)}) relative to those 
of the bulk generated fluctuations $C_b(x,\omega)$ (\ref{Cb(x,omega)}).

\subsection{Model for inlet fluctuations}\label{IVA}

In Ref. \cite{LR} $D(\omega)$ was estimated by projecting 
transverse momentum current fluctuations in thermal equilibrium onto the
critical mode of the 
hydrodynamic field equations. These fluctuations were assumed to be advected 
towards the inlet
from an upstream part $x<0$  without deterministic driving,
$\underline{\mu}=-1$, and to enter the main supercritical part of the
experimental cell , e.g., via a porous
plug at $x=0$.
Thus, the Rayleigh number Ra or the Taylor number T in the 
upstream part was taken to be zero in \cite{LR}. In experiments, however,
one often has an upstream section $-\underline{L}<x<0$ to the left of the
inlet at  $x=0$ in which the fluid is equilibrated at a finite subcritical
driving $\underline{\mu}<0$. In the remainder of this paper we mark
quantities characterizing such a subcritical upstream part of the system
by an underbar.

Babcock, Ahlers, and Cannell \cite{BAC} have numerically solved the stochastic
CGLE for such a setup with $\underline{L}=0.1,\underline{\mu}=-0.1$,
and boundary condition $A(x=-\underline{L},t)=0$.
We consider a situation where $\underline{L}$ is sufficiently large. Then
perturbations, e.g., from the pumping machinery that are swept at
$x=-\underline{L}$ into the subcritical upstream part of the apparatus have
decayed by the time the fluid reaches the inlet, $x=0$, of the supercritical
part. Thus, one ideally has at $x=0$ only fluctuations in the hydrodynamic
fields that are sustained by thermally activated fluctuating forces.
The most important field fluctuations are those with  wave numbers
close to the critical one and with spatial variation in the directions
perpendicular to the through-flow similar to the critical mode.
We approximately describe the dynamics of the amplitude of these modes 
with the linear stochastic CGLE (\ref{GLE_Ax}), now however for subcritical
driving $\underline{\mu}<0$. Therein,  the
spectrum of amplitude fluctuations at $x=0$ that are generated at subcritical
driving by thermal forcing $f(x \leq 0,t)$ alone --- $\underline{L}$ is 
large --- is given by (\ref{Cinf}) with $\mu=\underline{\mu}<0$ 
\begin{equation}
\label{Destimate}
D(\omega) = C_\infty (\omega;\underline{\mu})
=\frac{F}{2 (1+c_1^2)\xi_0^4 \mid \underline{K_1} - \underline{K_2} \mid^2 }
\left(\frac{1}{Im\underline{ K_1}} - \frac{1}{Im\underline{ K_2}} \right). 
\end{equation}
The total spectral power of these noise generated fluctuations of the amplitude
$A(x=0)$ of the critical mode at the inlet 
\begin{equation}
D(t=0) = \int_{- \infty}^\infty \frac{d \omega}{2 \pi} D(\omega)
=\frac{F}{4 \tau_0 \xi_0 \sqrt{- \underline{\mu}}} 
\end{equation}
increases with increasing driving in the subcritical upstream section and 
diverges in the absence of nonlinearities in (\ref{GLE_Ax}) at the critical
driving $\underline{\mu}=0$. Thus, the estimate (\ref{Destimate}) is only
realistic for a situation where the noise generated amplitudes in the upstream
part are small enough to neglect the nonlinear contributions in the
hydrodynamic field equations, i.e., for $\underline{\mu}$ not too close to
zero. If that is the case then a variation $\propto 1/\sqrt{- \underline{\mu}}$ 
in the inlet contribution should be experimentally observable.

In Fig. \ref{plotD(omega)} we show the inlet spectrum $D(\omega)$
(\ref{Destimate}) normalized by $\tau_0 D(t=0)$ versus reduced 
frequency $\omega \tau_0$ for several subcritical upstream control 
parameters as indicated. The inlet spectrum $D(\omega)$ is dominated by the
term $1/Im\underline{ K_1}$. It controls the $\underline{\mu}$ variation of
the peak height and of the peak width of $D(\omega)$ and it governs the 
strong enhancement of those modes around $\omega=0$ that can grow in the 
supercritical downstream part of the system.  

Finally we should like to comment on the
objection \cite{SBH} against using in a system without translational 
invariance thermal equilibrium spectra for inlet fluctuations that come from
the $\underline{\mu}=-1$ upstream part in \cite{LR}. This seems
to be well justified in view of the fact that downstream forces do not
influence upstream locations --- downstream generated perturbations cannot
propagate upstream if the through-flow is smaller than the convectively
unstable threshold. Thus for any $x$ in the upstream section that is
sufficiently far away from $\underline{L}$ one should observe fluctuations
as in an infinite system. For the stochastic CGLE this is reflected by
the equality of (\ref{Cx_inf}) and (\ref{Cinf(omega)}). But the corresponding
property holds also for the full linear field equations with through-flow.

\subsection{Comparison of inlet and bulk generated fluctuations}\label{IVB}

In Fig. \ref{CinCb} we show the downstream evolution of the spectral
contributions $C_{in}(x,\omega)$ (\ref{Cin(x,omega)}) and 
$C_b(x,\omega)$ (\ref{Cb(x,omega)}) from inlet and
bulk forcing, respectively, to the total spectrum of amplitude fluctuations.
The contributions are reduced by $F/\xi_0$ measuring the forcing strength. 
The inlet fluctuations $C(x=0,\omega)=C_{in}(x=0,\omega)=D(\omega)$ are taken
from the model of the previous subsection with a subcritical upstream
driving of $\underline{\mu}=-0.1$. The Reynolds number is Re=2 and the
supercritical downstream driving is $\mu=\mu^{c}_{conv}/2=0.03$.
For these parameters the band limits (\ref{omega_b}) 
of amplification $Im K_1(\omega) < 0$ are $\omega_-\tau_0 = - 0.084$ and
$\omega_+\tau_0  =  0.086$, respectively. While the logarithmic 
scale of
the ordinate tends to obscure the boundaries between growing and decaying
frequency modes the left border $\omega_-$ of the amplification band can be 
identified in the plots of $C_{in}(x,\omega)$  
of Fig. \ref{CinCb} by their crossing point ---  there $C_{in}$-spectra 
at different $x$ have the same value for $Im K_1= 0$ .

The bulk contribution $C_b$ which vanishes 
at the inlet is at a distance of one correlation length $\xi_0$ still smaller
than the inlet contribution. Then, with growing downstream distance
$C_b$ becomes larger. At $x=20\xi_0$ the bulk contribution has almost reached
its large-$x$ single exponential growth asymptotics
$\propto e^{-2 ImK_1(\omega)x}$. There the contribution to 
$C_b$ from the last term $\propto e^{i[K_1(\omega)-K_2(\omega)]x}$ in
(\ref{Cb(x,omega)}) is already small.

That can also be seen from Fig. \ref{Cx}. There we show for the same 
supercritical downstream driving $\mu=\mu^{c}_{conv}/2=0.03$ that
was used in Fig. \ref{CinCb} the downstream growth of the spectral peak
$C_b(x,\omega=0)$ in comparison with the peak height $C_{in}(x,\omega=0)$ of
the inlet contribution --- for our parameters the spectral maxima are located
practically at $\omega=0$. $C_{in}$ was evaluated for several subcritical 
upstream driving values of $\underline{\mu}$ as indicated. The dashed line 
labelled $C_{SBH}$ is discussed in Appendix \ref{appb3}. Thus, for the 
values of $\underline{\mu}$ in Fig. \ref{Cx} the bulk contribution rapidly
outgrows the inlet contribution. A similar observation was made also in the
numerical treatment of Babcock, Ahlers, and Cannell \cite{BAC}.
 
The slope of the straight lines $C_{in}$ and of the asymptotics of $C_b$ 
and $C_{SBH}$ is given by $-2 Im K_1^0$ where the superscript $0$ indicates
that the eigenvalue $K_1$ is taken at $\omega=0$.
In the far downstream region the ratio of the spectral maxima of $C_b$
and $C_{in}$ approaches according to (\ref{large_x}) the limiting value 
\begin{equation}
\lim_{x\to \infty}
\frac { C_b(x,\omega=0)}{C_{in}(x,\omega=0)} =
- \frac{F/D(\omega=0)}{2 (1+c_1^2) \xi_0^4 \mid K_1^{0*} - K_2^0 \mid^2 }  
\left(\frac{1}{Im K_1^0} + \frac{1}{Im K_2^0} \right) .
\label{Cb/Cin}
\end{equation}
If one uses for $D(\omega=0)$ the result (\ref{Destimate}) of our model  
one finds for the parameters of Fig.\ref{Cx} that the bulk generated
downstream fluctuations $C_b$ dominate over the fluctuations
that are swept into the system through the inlet whenever the upstream 
driving $\underline\mu<-0.025$. If the downstream
driving $\mu$ is smaller than in the example of Fig.\ref{Cx} then the size of 
inlet fluctuations that would ensure $C_{in}$ to be comparable to $C_b$ 
for large $x$ would have to be larger -- i.e., $\underline\mu$
would have to be even closer to zero.

Thus we conclude that {\em typically}
the large (small)-$x$ behavior is governed by bulk (inlet) fluctuations.
However, our upstream model clearly shows that in appropriately designed
experimental upstream sections --- with $\underline L$ being long enough 
to avoid influence of the pumping machinery operating at $x=-\underline L$ 
--- one can tune and enhance the inlet noise at $x=0$ almost
arbitrarily by selecting
the subcritical upstream driving $\underline\mu$ accordingly. Finally we
should like to stress that our model for inlet fluctuations contains only 
internal thermal fluctuations generated in the subcritical upstream part.
It does not contain other perturbations, e.g., from the pumping machinery 
that might in an experimental setup be advected towards the inlet. Their
contribution to the inlet fluctuation spectrum would have to
be added to our model spectrum (\ref{Destimate}). However, since our general
formulas in Sec. \ref{III} have been derived for arbitrary inlet
fluctuations with spectrum $D(\omega)$ they also apply to such a situation.


\section{Conclusion}\label{VI} 
 
We have investigated the 
effect of inlet fluctuations and bulk stochastic forcing on the noise 
sustained pattern growth in semiinfinite systems, $0 \leq x$, with 
through-flow in the convectively unstable parameter regime. We have 
used a 1D CGLE that is appropriate for forwards bifurcating 
structures in order to quantitatively compare both effects on the 
statistical dynamics of the downstream growth of the complex pattern 
amplitude $A (x,t)$.  
 
To that end we have solved the linear stochastic CGLE subject to 
arbitrary, statistically stationary boundary conditions at the inlet 
$x = 0$ in the presence of bulk additive forcing $f (x,t)$ with a spatial 
Laplace transform. This method is well suited 
for the semiinfinite geometry: inlet and bulk forcing are 
easily dealt with separately since they both appear explicitly and 
additively via boundary and inhomogeneous terms, respectively. 
Furthermore, this approach easily allows to identify the contributions 
to the general mathematical solution from a pole at $K_2(\omega)$ in  
the complex wave-number plane that have to be discarded 
on physical grounds since they would cause growth even for subcritical 
control parameters. The physically motivated restriction is equivalent 
to choosing the "space-retarded, causal" Green function that describes 
propagation away from (and not towards) a localized perturbation 
pulse.  The restriction imposes a relation between the inlet 
conditions on $A$ and $\partial_x A$ on the one hand and the bulk 
forcing $f$ on the other hand. Thus, for example the inlet derivative 
$\partial_x A$ is fixed in the physically relevant solution of the 
second order (in space) CGLE by $A$ at $x = 0$ and the bulk forcing 
$f$. 
 
We have investigated the effect of arbitrary statistically stationary inlet 
fluctuations that are independent from the bulk stochastic forces and 
that are, e. g., advected through the inlet at $x = 0$ into the system 
$x > 0$ with the through-flow from an upstream subsystem, $x < 0$. 
Then we have considered a simple model for such an upstream subsystem
that contains thermal fluctuations only: It is 
held at some subcritical driving $\underline{\mu} < 0$ and the 
thermally generated fluctuations that are advected towards the inlet 
are described by the stochastic CGLE. Thus, $\underline{\mu}$ allows to 
tune the strength of the advected inlet fluctuation spectrum.  
 
If the driving $\mu$ in the downstream part $x > 0$ is subcritical then 
all frequency modes are damped. Then the correlation spectrum $C (x, 
\omega)$ of amplitude fluctuations is at large downstream distances 
from the inlet no longer influenced by the inlet fluctuations at $x = 
0$ and $C (x,\omega)$ approaches the translational invariant spectrum 
of an infinite system. However, for supercritical downstream driving 
$\mu > 0$ inlet-forced as well as bulk-forced amplitude fluctuations 
are growing in downstream direction for frequencies $\omega$ within a 
finite band. Typically the contribution $C_b (x, \omega)$ caused 
by the additive bulk forces to the correlation spectrum $C (x, 
\omega)$ of amplitude fluctuations at the downstream location $x$ 
rapidly outgrows with increasing $x$ the contribution $C_{in} (x, 
\omega)$ that follows from the upstream model for inlet fluctuations. 
In the large-$x$ regime far from the inlet both contributions, $C_{in} 
(x, \omega)$ as well as $C_b (x, \omega)$, grow $\sim e^{- 2 \, Im \, 
K_1 (\omega) x}$ with $- Im \, K_1 \, > \, 0$ having a maximum close 
to $\omega = 0$. But the prefactors in $C_b$ and $C_{in}$ are such 
that there the amplitude fluctuation spectrum $C (x, \omega)$ 
typically is dominated by the bulk contribution $C_b (x, \omega)$. 
However, we have shown with our model how to realize experimental  
setups with subcritical upstream driving where the inlet  
fluctuations could be well controlled and well defined --- 
uninfluenced by e.g. the pumping machinery --- and tuned to be 
large enough to dominate over the bulk part $C_b$.  
 
Finally, a detailed comparison with a recent approach of  
SBH showed that the boundary condition that they use at the 
inlet implies fluctuations there which are strongly correlated  
with the bulk forcing. This enhances the amplitude fluctuations of the 
growing modes and thereby tends to overestimate the effect of thermal bulk 
forcing.  
\acknowledgments
This work was supported by the Go West program of the EC and by the DAAD.
One of us (AS) acknowledges the hospitality of the Universit\"{a}t des
Saarlandes. We thank A.~Recktenwald and P.~B\"{u}chel for help with the
figures.

\appendix

\section{ Green function}\label{appa}

We are interested in the Green function
$G(x,x_0;t-t_0)$ of the CGLE as a solution of 
\begin{equation}
\left[ \tau_0 (\partial_t + v \partial_x) - (1+ic_0)\mu 
- (1+ic_1)\xi_0^2\partial_x^2\right] G(x,x_0;t-t_0) =
\delta(x-x_0) \delta(t-t_0)
\label{Green_GLE}
\end{equation}
subject to an arbitrary boundary condition $G(x=0,x_0;t-t_0)$
at the inlet. Because of the semiinfinite geometry  $G$ depends on
$x$ and the location $x_0$ of the $\delta$-pulse separately.
The temporal Fourier transform of $G$ 
\begin{equation}  
G(x,x_0;\omega) = \int_{-\infty}^{\infty} dt e^{i\omega (t-t_0)} 
G(x,x_0;t-t_0)  
\end{equation}
is conveniently written as the sum of two contributions
\begin{equation}
G(x,x_0;\omega) = 
e^{iK_1x} G(x=0,x_0;\omega) +\tilde{G}(x,x_0;\omega).
\label{GF_sum} 
\end{equation}
Then the first term propagates the inlet condition in downstream direction
and the boundary condition on the bulk contribution
\begin{eqnarray}
\tilde{G}(x,x_0;\omega) \nonumber\\ 
= \frac{i}{(1+ic_1)\xi_0^2 (K_1 - K_2)}
\left[\theta(x-x_0) e^{i K_1 (x-x_0)} + \theta(x_0-x) e^{i K_2 (x-x_0)} 
- e^{i ( K_1 x - K_2 x_0)} \right] 
\end{eqnarray}
 is $\tilde{G} (x=0,x_0;\omega) = 0$.
The physical solution (\ref{A(K,omega)restricted}) of the linear stochastic
CGLE can then be written in the form
\begin{equation} 
A(x,\omega) = \int_{0}^{\infty} dx_0
G(x,x_0;\omega)f(x_0,\omega) 
\end{equation}
and 

\begin{equation}
A(x,t) = \int_{0}^{\infty} dx_0\int_{-\infty}^{t} dt_0
G(x,x_0;t-t_0) f(x_0,t_0).
\end{equation}
The $x_0$-integration covers the whole spatial domain $x_0 \geq 0$. Causality
implies contributions to come only from times $-\infty < t_0 \leq t$ prior to
the observation time $t$. The inlet conditions on $G$ is defined by the 
requirement that for any given $f(x_0,\omega)$ the integral
\begin{equation}
\int_0^\infty \,dx_0 G(x=0,x_0;\omega)f(x_0,\omega) = A(x=0,\omega)
\end{equation}
yields the {\em imposed} inlet condition $ A(x=0,\omega)$ that
can be chosen arbitrarily and in particular independent of $f$.

For the sake of comparison with the Green function $G_{SBH}$
used by SBH (cf. Appendix \ref{appb1}) we mention that $G$ (\ref{GF_sum})
can be rewritten as the following decomposition
\begin{equation}
\label{GFdecompSBH}
G(x,x_0;\omega) 
= e^{iK_1x} \left[G(x=0,x_0;\omega) - G_{SBH}(-x_0;\omega) \right]
+G_{SBH}(x-x_0;\omega).
\end{equation}
Here 
\begin{eqnarray}
\label{GSBH(x,omega)}
G_{SBH}(x-x_0;\omega) 
= \frac{i}{(1+ic_1)\xi_0^2 (K_1 - K_2)} \left[\theta(x-x_0) 
e^{i K_1 (x-x_0)} + \theta(x_0-x) e^{i K_2 (x-x_0)} \right] 
\end{eqnarray}
is the Green function used by SBH and
\begin{eqnarray}
G_{SBH}(-x_0;\omega) 
= \frac{i}{(1+ic_1)\xi_0^2 (K_1 - K_2)} e^{-i K_2 x_0} 
\end{eqnarray}
is its value at the inlet. Note that the "space-retarded, causal" nature
of the Green function can readily be read off from (\ref{GSBH(x,omega)}).
It describes spreading off ---not contraction towards ---
the perturbation pulse at $x_0$: the first (second) exponential in
(\ref{GSBH(x,omega)}) accounts for the spreading in downstream (upstream)
direction with complex wave number $K_1$ ($K_2$). Downstream growth occurs
for frequencies with $Im K_1(\omega) < 0$ while the spreading in upstream 
direction is always damped given that $Im K_2 < 0$.

Since the linear operator appearing in the
CGLE (\ref{GLE}), i.e., the square bracket in (\ref{Green_GLE}) is 
$\propto (\partial_x - i K_1)(\partial_x - i K_2)$ it is clear from
(\ref{GFdecompSBH}) that both, $G$ and $G_{SBH}$, solve
Eq.(\ref{Green_GLE}), however with different inlet conditions. 


\section{Comparison with SBH}\label{appb}
\subsection{ Green function $G_{SBH}(x-x_0;\omega)$}\label{appb1}
Here we derive the expression (\ref{GSBH(x,omega)}) for  
\begin{equation}
\label{GSBH(y,omega)}
G_{SBH}(y;\omega) 
= \frac{i}{(1+ic_1)\xi_0^2 (K_1 - K_2)} \left[\theta(y) 
e^{i K_1 y} + \theta(-y) e^{i K_2 y} \right] 
\end{equation}
with $y=x-x_0$ by temporally Fourier transforming the Green function
\begin{eqnarray}
\label{GSBH(y,tau)}
G_{SBH}(y;\tau) = \frac{\theta(\tau)}{\tau_0} e^
{(1+ic_0)\mu \tau/\tau_0}\int_{-\infty}^\infty \frac{dk}{2 \pi} 
\exp\left[-(1+ic_1)\xi_0^2 
k^2 \frac {\tau}{\tau_0}+ik(y-v\tau)\right] 
\label{G_kintegr} 
\end{eqnarray} 
given in eqs. (C18, C19) of SBH for $\tau=t-t_0>0$. Inverting the sequence
of integrations one obtains 
\begin{eqnarray}
G_{SBH}(y;\omega) &=& \frac{1}{\tau_0} \int_{-\infty}^\infty
\frac{dk}{2\pi}e^{iky} \int_{0}^\infty d\tau\exp\left[i\omega\tau + 
(1+ic_0)\mu\frac{\tau}{\tau_0} - (1+ic_1)\xi_0^2 k^2\frac{\tau}{\tau_0} - 
ikv\tau \right]
\nonumber\\
&=& \frac{1}{\tau_0} \int_{-\infty}^\infty
\frac{dk}{2\pi}e^{iky} \int_{0}^\infty d\tau
\exp\left[-(1+ic_1)\xi_0^2
\left(k-K_1\right) \left(k-K_2\right) \frac{\tau}{\tau_0} \right] 
 \nonumber\\
&=& \frac{1}{(1+ic_1)\xi_0^2} \int_{-\infty}^
\infty \frac{dk}{2\pi} \frac{e^{iky}}{(k-K_1)(k-K_2)}. 
\label{Green_SBH2}
\end{eqnarray} 
The above integral along the real $k$-axis can be calculated as a contour
integral in the complex $k$-plane with the method of residues. For 
$y>0$ ($y<0$) one has to close the integration contour in the upper (lower)
complex halfplane. 

The pole at $K_2$ contributes only for $y<0$
since it lies with $Im K_2<0$ always in the lower half of the complex
$k$-plane. It yields the second term in (\ref{GSBH(y,omega)}). The pole at 
$K_1$  lies for subcritical $\mu<0$ in the upper  half of the complex
$k$-plane. It causes the term $\propto e^{iK_1y}$ for $y>0$ in
(\ref{GSBH(y,omega)}) coming from the contour closed in the upper half of the
complex $k$-plane. However, when $\mu>0$ this pole crosses the real $k$-axis
and moves into the lower $k$-plane for frequencies within the band 
(\ref{omega_b}) of growing modes. In order to retain its contribution in 
(\ref{GSBH(y,omega)}) for $y>0$ one has to
deform the original integration path along the real $k$-axis into a contour
reaching into the lower $k$-plane that encloses $K_1$ such that this pole
remains in the interior.

Alternatively 
one can first perform the k-integration in (\ref{GSBH(y,tau)})
and then Fourier transform the result
\begin{eqnarray}
G_{SBH}(y;\omega)
= \frac{1}{2 \xi_0 \sqrt{(1+ic_1)\pi \tau_0}}
\int_{0}^{\infty}\frac{d\tau}{\sqrt{\tau}} 
\exp\left[i\omega\tau + (1+ic_0)\mu\frac{\tau}
{\tau_0} - \frac{(y-v\tau)^2 \tau_0}{4(1+ic_1)\xi_0^2 \tau } \right].
\end{eqnarray}
Using $\sqrt{\tau}$ as integration variable one finds with Eq. (3.472-1)
of Gradshteyn and Ryzhik \cite{RG-IntTab} that
\begin{mathletters}
 \label{Green_SBHalt}
\begin{equation}
G_{SBH}(y;\omega)=\frac{1}{\gamma}e^{\alpha y -\beta \sqrt{y^2}}
\end{equation}
with 
\begin{eqnarray}
\alpha  &=& \frac{v\tau_0}{2(1+ic_1)\xi_0^2 }
\\
\beta  &=& \sqrt{\alpha^2 - 
\frac{i\omega\tau_0 + (1+ic_0)\mu}{(1+ic_1)\xi_0^2}}
\\
\gamma  &=& 2\beta(1+ic_1)\xi_0^2.
\end{eqnarray}
\end{mathletters} 
Expressing $\alpha, \beta$, and $\gamma $ in terms of the eigenvalues 
$K_{1,2}$~(\ref{eigenvalues}) and using the fact that 
$\sqrt{y^2}/y = sign( y) $ one verifies that (\ref{Green_SBHalt}) agrees with
(\ref{GSBH(y,omega)}).

\subsection{ Amplitude fluctuations $A_{SBH}(x,\omega)$}\label{appb2}

The amplitude fluctuations of SBH are expressed with the help of the Green 
function $G_{SBH}$ as 
\begin{equation}
A_{SBH}(x,t) = \int_{0}^{\infty} dx_0\int_{-\infty}^{t} dt_0
G_{SBH}(x-x_0;t-t_0) f(x_0,t_0).
\label{ASBH(x,t)}
\end{equation}
Here we deal with the statistically stationary situation where the stochastic
forces have been operating since the time $t_0=-\infty$. On the other hand,
SBH enforce the initial condition $A_{SBH}(x,t=0)=0$ at time $t=0$ by 
integrating in (\ref{ASBH(x,t)}) only from $t_0=0$ to $t_0=t$ --- cf. Eq. (C.20)
in \cite{SBH}. Since we are interested however in comparing the {\em longtime}
properties of their results with ours this difference does not play a role ---
the formulas labelled with the subscript SBH in this and the next subsection
refer to the longtime limit of the SBH results in \cite{SBH}.

In frequency space the longtime amplitude fluctuations of SBH 
follow with (\ref{GSBH(y,omega)}) to be 
\begin{eqnarray}
\label{ASBH(x,omega)}
A_{SBH}(x,\omega) = \frac{i}{(1+ic_1)\xi_0^2 (K_1 - K_2)}
\Biggl[ \int_0^x  dx_0 e^{i K_1 (x-x_0)} f(x_0,\omega) \nonumber\\
+ \int_x^\infty dx_0 e^{i K_2 (x-x_0)} f(x_0,\omega) \Biggr].
\end{eqnarray}
Note that the SBH solution is a very special one in the sense that
it assumes the longtime amplitude fluctuations at the inlet to be of the form
\begin{eqnarray}
\label{ASBH(x=0,omega)}
A_{SBH}(x=0,\omega)
& = & \frac{i}{(1+ic_1)\xi_0^2 (K_1 - K_2)}
\int_{0}^{\infty} dx_0 e^{-iK_2 x_0} f(x_0,\omega)
\nonumber\\
& = & \frac{-f(K_2,\omega)}{(1+ic_1)\xi_0^2 (K_1 - K_2)}.
\end{eqnarray}
Thus, the inlet boundary condition is not treated as an independent quantity
that can be imposed on the solution of the CGLE. Rather the 
inlet boundary value for $A$ is related to and determined explicitly by
the bulk forcing $f(x_0,\omega)$. However, when we impose in our {\em general}
solution (\ref{A(x,omega)restricted}) the special inlet condition
$A_{SBH}(x=0,\omega)$ of SBH do we get the same result as SBH. In fact, our 
general solution (\ref{A(x,omega)restricted}) can be written in the form
\begin{equation}
A (x,\omega) = A_{SBH}(x,\omega) + 
e^{i K_1 x} \left[A (x=0,\omega) - A_{SBH}(x=0,\omega)  \right]
\label{A_equiv} 
\end{equation}
with an inlet  boundary condition $A (x=0,\omega)$ that is still free to be 
chosen. 

Finally we mention that the derivative of the special SBH solution at the inlet
\begin{eqnarray}
\partial_x A_{SBH}(x,\omega) \mid_{x=0}
& = &\frac{- K_2}{(1+ic_1) 
\xi_0^2 (K_1 - K_2)} \int_0^\infty dx_0 e^{-i K_2 x_0} f(x_0,\omega)
\nonumber\\
& = & iK_2 A_{SBH}(x=0,\omega)
\end{eqnarray}
is such that also the SBH solution obeys the relation (\ref{restriction})
\begin{equation}
i \left( \partial_x - i K_1 \right) A_{SBH}(x, \omega)\mid_{x = 0}
+ \frac{f (K_2, \omega)}{(1+ic_1)\xi_0^2}=0.
\end{equation}
discussed in Sec. \ref{IIIA} which prevents unphysical behavior for
$x \to \infty$.

\subsection{ Correlation function $C_{SBH}(x,\omega)$}\label{appb3}

Using $< A_{SBH}(x,\omega) \left[ A_{SBH} (x,\omega') \right]^* > = 
2\pi\delta(\omega-\omega') C_{SBH}(x,\omega)$ to evaluate the correlation
spectrum one obtains from (\ref{ASBH(x,omega)})
\begin{equation}
\label{CSBH(x,omega)}
C_{SBH}(x,\omega) = 
\frac{F}{2 (1+c_1^2)\xi_0^4  \mid K_1 - K_2 \mid^2 } \left[\frac{1}{Im K_1}
\left(1 - e^{-2Im K_1 x} \right) - \frac{1}{Im K_2} \right] 
\end{equation}
with inlet correlations being given by 
\begin{equation}
\label{CSBH(x=0,omega)}
C_{SBH}(x=0,\omega) =  
\frac{F}{2 (1+c_1^2)\xi_0^4  \mid K_1 - K_2 \mid^2 } \frac{-1}{Im K_2}.
\end{equation}
These correlation spectra have to be compared with our result
(\ref{C(x,omega)}) and (\ref{D(omega)}). Note that $F = 2\xi_0\tau_0 F_A^{SBH}$
as mentioned already in Sec. \ref{IIC}.

To that end we show in Fig.\ref{CandC_SBH} amplitude fluctuation spectra at
the inlet, $x=0$, and further downstream, $x=10\xi_0$, for Re=2 and 
$\mu=\mu^{c}_{conv}/2=0.03$.  Full lines show our
results using inlet fluctuation spectra (Sec. \ref{IVA}) obtained for different
subcritical upstream driving $\underline{\mu}$ as indicated.
The most important difference
is that in SBH the inlet fluctuations are determined by the bulk forcing $f$
with a spectrum (\ref{CSBH(x=0,omega)}) that is fixed by $F, Re$, and $\mu$
whereas in our more general
case the inlet boundary condition on the amplitude $A$ is still open.
In the case that upstream fluctuations outside the main system generate 
our $A(x=0,\omega)$ the latter is statistically independent of amplitude 
fluctuations $A_b (x,\omega)$ (\ref{Ab(x,omega)}) that are generated in the
bulk by the random forces $f$, i.e.,
$< A(x=0,\omega) \left[ A_b (x,\omega') \right]^* > = 0$ (\ref{<A(x=0)Ab(x)>}).

This different correlation behavior influences not only the amplitude
fluctuations close to the inlet but also the large-$x$ correlation
spectrum of the growing frequency modes with $Im K_1<0$. Only the
correlation spectra of the {\em damped} modes with $Im K_1>0$ are in both 
approaches
the same for large $x$: $ C_{SBH}(x\to \infty,\omega) = C(x\to \infty,\omega)
 = C_\infty (\omega) $. (\ref{Cinf}) holds for the damped frequencies with
$Im K_1>0$ since at $x\to \infty$ these modes do no longer feel any 
influence of the inlet boundary condition. However, for the {\em growing}
modes with $Im K_1<0$ 
\begin{equation}
\label{CSBH(xtoinfty,omega)}
C_{SBH}(x\to \infty,\omega) \to 
-\frac{F}{2(1+c_1^2)\xi_0^4  \mid K_1 - K_2 \mid^2 } \frac{1}{Im K_1}
 e^{-2 Im K_1 x} 
\end{equation}
differs from 
\begin{equation}
\label{C(xtoinfty,omega)}
 C(x\to \infty,\omega) \to    
\left[ D(\omega)
- \frac{F}{2 (1+c_1^2) \xi_0^4 \mid K_1^* - K_2 \mid^2 }
\left(\frac{1}{Im K_1} + \frac{1}{Im K_2} \right) \right]e^{-2 Im K_1 x} .
\end{equation}
Thus, if one just replaces in (\ref{C(xtoinfty,omega)}) our inlet 
spectrum $D(\omega)$ by
$C_{SBH}(x=0,\omega)$ (\ref{CSBH(x=0,omega)}) one does not recover 
the SBH result (\ref{CSBH(xtoinfty,omega)}) since the SBH spectrum was 
derived for inlet fluctuations
$A_{SBH}(x=0,\omega)$ (\ref{ASBH(x=0,omega)}) which are correlated with the
bulk forcing $f$ while our inlet boundary condition on  
$A$ is chosen to be externally determined and thus independent
of the bulk forcing.

These additional correlations between inlet,
$x=0$, and downstream positions, $x>0$, that are present in the SBH solution
enhance the SBH spectra compared to our spectra. Consider, e.g., the case 
$\underline{\mu}=-0.05$ in Fig.\ref{CandC_SBH} for which our inlet fluctuation
spectrum is even larger than $C_{SBH}(x=0,\omega)$. Nevertheless, the
downstream amplitude fluctuation spectrum $C_{SBH}(x,\omega)$ rapidly
outgrows our $C(x,\omega)$. The effect of this correlation enhanced downstream
growth of SBH can also be seen in Fig.\ref{Cx} --- compare the dashed line
for the peak height $C_{SBH}(x,\omega=0)$ with our result $C(x,\omega=0)$
for $\underline{\mu}=-0.05$. 
  
This effect can be measured quantitatively, e.g., by comparing the prefactors
of the large-$x$ exponential growth behaviour of 
(\ref{CSBH(xtoinfty,omega)}) with that of (\ref{C(xtoinfty,omega)}). 
To that end we ignore the imaginary coefficients
$c_0$ and $c_1$ in view of their smallness \cite{parameters}. Then the ratio
of the spectral peak heights at $\omega=0$ has the form
\begin{equation}
\label{ratio}
\frac{C_{SBH}(x\to \infty,\omega=0)}{C(x\to \infty,\omega=0)} \to
\frac{\hat\mu}{2(1-\hat\mu)\left(1-\sqrt{1-\hat\mu}\right)} 
\enspace  \frac{\sqrt{1-\underline{\hat\mu}}} {\sqrt{1-\underline{\hat\mu}}
 - \hat\mu/\underline{\hat\mu}},
\end{equation}
where 
\begin{equation}
\hat\mu = \mu/\mu^c_{conv}>0, \qquad \hat{\underline{\mu}} =
{\underline\mu}/\mu^c_{conv}<0
\end{equation}
denote reduced control parameters. For not too large $\hat\mu$ the ratio
(\ref{ratio}) is close to 1. However, when $\hat\mu$ approaches 1 it becomes 
large. The coefficients $c_0, c_1$ remove the singularity present at 
$\hat\mu=1$ in (\ref{ratio}). 
In Fig.~5 the ratio $C_{SBH}(x,\omega=0)/C(x,\omega=0)$ of the peak heigths 
of the {\em full}
correlation spectra is shown at $x=20\xi_0$ versus reduced downstream driving
$\hat\mu$. This ratio differs only slightly from the large-$x$ asymptote
(\ref{ratio}) because of a small contribution from the last term in  
(\ref{Cb(x,omega)}). 

The inlet fluctuations resulting from the upstream part
of the system would have to be very large, i.e., $\underline{\mu}$ would 
have to be very close to zero in order to cause downstream
fluctuation amplitudes that are as large as those of the SBH approach.
Furthermore, if thermal equilibrium fluctuations were causing the inlet
fluctuations as described by the unforced case $\underline{\mu}=-1$ in our 
upstream model the discrepancy would be even larger. 
Thus, in agreement with an 
assessment made in the introduction of \cite{SBH} we conclude that the SBH
result overestimates the effect of thermal noise mainly because the
inlet boundary is not treated adequately.
 

%
%

\begin{figure} \caption[]
{ Spectrum $D(\omega)$ (\ref{Destimate}) of amplitude fluctuations
at the inlet normalized by $\tau_0 D(t=0)$ versus reduced 
frequency $\omega \tau_0$ for several $\underline{\mu}$. Parameters 
 \cite{parameters} are for Re=2.}
\label{plotD(omega)}
\end{figure}

\begin{figure} \caption[]
{Downstream evolution of the reduced contributions 
$C_{in}(x,\omega)$ (\ref{Cin(x,omega)}) and 
$C_b(x,\omega)$ (\ref{Cb(x,omega)}) to amplitude fluctuation spectra. Being 
practically symmetric in $\omega$ for our parameters
$C_{in}\xi_0/F$ (left side) and $C_b\xi_0/F$ (right side) can be
continued symmetrically to the other side.
Inlet fluctuations
 $C_{in}(x=0,\omega)=D(\omega)$ are taken
from our model with subcritical upstream
driving $\underline{\mu}=-0.1$. Parameters \cite{parameters} are Re=2,
$\mu=\mu^{c}_{conv}/2=0.03$.}
\label{CinCb}
\end{figure}

\begin{figure} \caption[]
{Downstream growth of spectral peaks. They are located very close to
$\omega=0$ for our parameters \cite{parameters} Re=2,
$\mu=\mu^{c}_{conv}/2=0.03$. Thin lines show $C_{in}(x,\omega=0)\xi_0/F$
coming from inlet fluctuations (Sec. \ref{IVA}) obtained for different
subcritical upstream driving $\underline{\mu}$. For the  SBH result 
$C_{SBH}(x,\omega=0)\xi_0/F$ (\ref{CSBH(x,omega)}) 
represented by the thick dashed line the inlet fluctuations are fixed
by construction via $\mu, Re$ (cf. Appendix \ref{appb3}).}
\label{Cx}
\end{figure}

\begin{figure} \caption[]
{Reduced correlation spectra of amplitude fluctuations at (a) inlet,
$x=0$, and (b) further downstream, $x=10\xi_0$. Parameters 
\cite{parameters} are Re=2, $\mu=\mu^{c}_{conv}/2=0.03$. Full lines show our
results using inlet fluctuation spectra (Sec. \ref{IVA}) obtained for different
subcritical upstream driving $\underline{\mu}$. Inlet fluctuations entering 
into the SBH result (dashed lines) are fixed
by construction via $\mu, Re$ (cf. Appendix \ref{appb3}).}   
\label{CandC_SBH}
\end{figure}

\begin{figure} \caption[]
{Ratio of spectral peak heights of amplitude fluctuations according to
SBH and to our work. The correlation spectra have almost reached their
large-$x$ asymptotic form at $x=20\xi_0$. Our spectra were evaluated
using inlet fluctuations (Sec. \ref{IVA}) obtained for different
subcritical upstream driving $\underline{\mu}$. Inlet fluctuations entering 
into the SBH result are fixed
by construction via $\mu, Re$ (cf. Appendix \ref{appb3}).
Parameters \cite{parameters} are Re=2, $\mu=\mu^{c}_{conv}/2=0.03$.}  
\label{C_SBH/C}
\end{figure}

\end{document}